\documentclass[10pt,journal,compsoc]{IEEEtran}

\usepackage{tabularx}
\usepackage{graphicx}
\usepackage[T1]{fontenc}

\usepackage[ruled,vlined,linesnumbered]{algorithm2e}
\SetAlFnt{\footnotesize}\SetAlCapFnt{\footnotesize}\SetAlCapNameFnt{\footnotesize}

\SetCommentSty{mycommfont}

\ifCLASSOPTIONcompsoc
  \usepackage[nocompress]{cite}
\else
  \usepackage{cite}
\fi
\ifCLASSINFOpdf
\else
\fi
\usepackage{amsmath}
\usepackage[cmintegrals]{newtxmath}
\DeclareMathOperator{\argmin}{argmin}

\usepackage[bbgreekl]{mathbbol}
\usepackage{amsfonts}

\DeclareSymbolFontAlphabet{\mathbb}{AMSb}
\DeclareSymbolFontAlphabet{\mathbbl}{bbold}

\usepackage{url,multirow,morefloats,floatflt,cancel,tfrupee}
\makeatletter

\AtBeginDocument{\@ifpackageloaded{textcomp}{}{\usepackage{textcomp}}}
\makeatother
\usepackage{colortbl}
\usepackage{xcolor}
\usepackage{pifont}
\usepackage[nointegrals]{wasysym}
\makeatletter

\def\mcWidth#1{\csname TY@F#1\endcsname+\tabcolsep}

\def\cAlignHack{\rightskip\@flushglue\leftskip\@flushglue\parindent\z@\parfillskip\z@skip}
\def\rAlignHack{\rightskip\z@skip\leftskip\@flushglue \parindent\z@\parfillskip\z@skip}

\usepackage{ifxetex}
\ifxetex\else\if@twocolumn\usepackage{dblfloatfix}\fi\fi

\AtBeginDocument{
\expandafter\ifx\csname eqalign\endcsname\relax
\def\eqalign#1{\null\vcenter{\def\\{\cr}\openup\jot\m@th
  \ialign{\strut$\displaystyle{##}$\hfil&$\displaystyle{{}##}$\hfil
      \crcr#1\crcr}}\,}
\fi
}

\AtBeginDocument{%
  \@ifpackageloaded{endfloat}%
   {\renewcommand\efloat@iwrite[1]{\immediate\expandafter\protected@write\csname efloat@post#1\endcsname{}}}{}%
}%

\def\BreakURLText#1{\@tfor\brk@tempa:=#1\do{\brk@tempa\hskip0pt}}
\let\lt=<
\let\gt=>
\def\processVert{\ifmmode|\else\textbar\fi}

\@ifundefined{subparagraph}{
\def\subparagraph{\@startsection{paragraph}{5}{2\parindent}{0ex plus 0.1ex minus 0.1ex}%
{0ex}{\normalfont\small\itshape}}%
}{}

\newcommand\role[1]{\unskip}
\newcommand\aucollab[1]{\unskip}
  
\@ifundefined{tsGraphicsScaleX}{\gdef\tsGraphicsScaleX{1}}{}
\@ifundefined{tsGraphicsScaleY}{\gdef\tsGraphicsScaleY{.9}}{}
\def\checkGraphicsWidth{\ifdim\Gin@nat@width>\linewidth
	\tsGraphicsScaleX\linewidth\else\Gin@nat@width\fi}

\def\checkGraphicsHeight{\ifdim\Gin@nat@height>.9\textheight
	\tsGraphicsScaleY\textheight\else\Gin@nat@height\fi}

\def\fixFloatSize#1{}
\let\ts@includegraphics\includegraphics

\def\inlinegraphic[#1]#2{{\edef\@tempa{#1}\edef\baseline@shift{\ifx\@tempa\@empty0\else#1\fi}\edef\tempZ{\the\numexpr(\numexpr(\baseline@shift*\f@size/100))}\protect\raisebox{\tempZ pt}{\ts@includegraphics{#2}}}}

\AtBeginDocument{\def\includegraphics{\@ifnextchar[{\ts@includegraphics}{\ts@includegraphics[width=\checkGraphicsWidth,height=\checkGraphicsHeight,keepaspectratio]}}}

\def\URL#1#2{\@ifundefined{href}{#2}{\href{#1}{#2}}}

\def\UrlOrds{\do\*\do\-\do\~\do\'\do\"\do\-}%
\g@addto@macro{\UrlBreaks}{\UrlOrds}

\@ifundefined{quoteAttrib}
	{}
	{}

\@ifundefined{titlequoteAttrib}
	{}{}

\newenvironment{title-quote}
	{\list{}{\fontsize{10pt}{12pt}\selectfont\leftmargin.5in\itshape\rightmargin\leftmargin}%
  \item\relax}
  {\endlist}

\makeatother

\makeatletter
\AtBeginDocument{\@ifpackageloaded{longtable}{%
\def\LT@makecaption#1#2#3{%
  \LT@mcol\LT@cols c{\hbox to\z@{\hss\parbox[t]\LTcapwidth{%
    \sbox\@tempboxa{#1{#2: } #3}%
    \ifdim\wd\@tempboxa>\hsize
      #1{#2: }\textsc{#3}%
    \else
      \hbox to\hsize{\hfil\box\@tempboxa\hfil}%
    \fi
    \endgraf\vskip\baselineskip}%
  \hss}}}
}{}}
\makeatother

   \makeatletter
  \def\fig@textbf{\textbf}
   \AtBeginDocument{\renewcommand\floatc@plain[2]{\setbox\@tempboxa\hbox{{\footnotesize#1.}\footnotesize\hskip.5em#2}%
    \ifdim\wd\@tempboxa>\hsize {\fig@textbf{\footnotesize#1.}}\footnotesize\hskip.5em#2\par
        \else\hbox to\hsize{\hfil\box\@tempboxa\hfil}\fi}}
    \makeatother
  
\usepackage{float}

\newfloat{scheme}{tbp}{Scheme}
\floatname{scheme}{Scheme}
\makeatletter
\@ifpackageloaded{endfloat}{\@ifpackagewith{endfloat}{figuresonly}{\DeclareDelayedFloat{scheme}{Scheme}}{\@ifpackagewith{endfloat}{tablesonly}{}{\DeclareDelayedFloat{scheme}{Scheme}}}}{}\makeatother

\begin{document}

%

\title{A Grid-based Approach for Convexity Analysis of a Density-based Cluster}
      

%
%
\author{Sayyed-Ahmad~Naghavi-Nozad\IEEEcompsocitemizethanks{\IEEEcompsocthanksitem Sayyed-Ahmad~Naghavi-Nozad is with Computer Engineering Department, AmirKabir University of Technology (Tehran Polytechnic), 424 Hafez Ave, Tehran, 15875-4413, Tehran, Iran, Tel.:~+98 (21) 64540, Email: sa\_na33@aut.ac.ir\newline\indent Sayyed-Ahmad~Naghavi-Nozad is the corresponding author.}}
\IEEEtitleabstractindextext{
\begin{abstract}
	This paper presents a novel geometrical approach to investigate the convexity of a density-based cluster. Our approach is grid-based and we are about to calibrate the value space of the cluster. However, the cluster objects are coming from an infinite distribution, their number is finite, and thus, the regarding shape will not be sharp. Therefore, we establish the precision of the grid properly in a way that, the reliable approximate boundaries of the cluster are founded. After that, regarding the simple notion of convex sets and midpoint convexity, we investigate whether or not the density-based cluster is convex. Moreover, our experiments on synthetic datasets demonstrate the desirable performance of our method.
\end{abstract}
    

}

\maketitle
      
\IEEEdisplaynontitleabstractindextext

%
\IEEEpeerreviewmaketitle

\section{Introduction}\label{sec_Intro}Convexity analysis of a set of points is an essential problem in computational geometry, as well as many other scientific fields. In the literature, there are some methods for calibrating the value space of the problem, like convex hull \unskip~\cite{berg2008computational}, voronoi diagram \unskip~\cite{aurenhammer1991voronoi} and delaunay triangulation \unskip~\cite{delaunay1934sphere}. In convex hull, we are trying to find the smallest convex set that contains the cluster objects. For instance, for cluster $ \mathcal{X} $ in two dimensions, the convex hull may be visualized as the shape enclosed by a rubber band stretched around $ \mathcal{X} $. In voronoi diagram, we calibrate the value space by partitioning it into regions, based on distance to points in a specific subset of the plane. Also, a delaunay triangulation for a given set of points is a triangulation, such that no point will be inside the circumcircle of any triangle of related points \unskip~\cite{fisher2004visualizing}. But the problem is that the essence of such approaches is about a set of points, and none of them have been employed to analyze the convexity of a cluster. Therefore, we are about to adopt a new method from such geometrical approaches, which can discuss on convexity of a density-based cluster.

In this paper, we propose a novel grid-based approach, which is both easy-to-understand and easy-to-implement, for finding the extreme marginal points of a dense cluster. These points include not only the outer margin of the cluster, but even the inner margins of it will be detected. After gaining such points, which constitute the approximate frontiers of the cluster, we can assume the cluster as a multidimensional shape, whose boundaries are calibrated. Then by utilizing the concept of the midpoint convexity\unskip~\cite{boyd2004convex} on these boundary points, we can evaluate the convexity of the cluster.

The paper is organized as follows: In Section~\ref{sec_BascConcp}, we present the basic concepts required to understand the fundamentals of the proposed method. In Section~\ref{sec_PropMeth}, the detailed descriptions of the proposed approach along with the novel algorithms are provided. In Section~\ref{sec_Expr}, experimental analyses are reviewed. Finally, we conclude the paper in Section~\ref{sec_Conc}.

\section{Basic Concepts}\label{sec_BascConcp}A number of definitions related to grid-based analysis of a cluster are reviewed. First of all, to find out that which points of the cluster are at its boundaries, we need to calibrate the space, which is covering the cluster, and it is like rasterizing a two-dimensional image, as then we know the exact position of every cluster point. Thus, we are utilizing a grid for this purpose. The major notations are represented in table~\ref{table-majNot}.

\begin{table}[!t]
	\renewcommand{\arraystretch}{1.15}
	\caption{Major Notations}
	\label{table-majNot}
	\scriptsize
	\centering
	\setlength{\tabcolsep}{.9pt}
	
	\begin{tabular}{p{1.5cm}p{6.5cm}}
		\hline
		\bfseries Notation & \bfseries Description\\
		\hline
		$ \mathcal{G} $ & The grid structure covering the value space\\
		$ \varepsilon $ & The grid accuracy\\
		$ t $ & Number of grid points in $ \mathcal{G} $\\
		$ \eta $ & Random sampling rate for grid points\\
		$ \mathcal{G}_s $ & Set of sampled grid points\\
		$ \mathbb{N}\left( g\right) $ & The neighboring set of grid point $ g $\\
		$ \left[ \mathcal{X} \right]_{n\times p} $ & Input cluster $ X $ with $ n $ objects and $ p $ attributes\\
		$ \mathcal{N}_\varepsilon \left( x \right) $ & The $ \varepsilon $ neighborhood of point $ x $\\
		$ D \left( x,y \right) $ & Euclidean distance between points $ x $ and $ y $\\
		$ M \left( x,y \right) $ & Midpoint of points $ x $ and $ y $\\
		$ \left| \cdot\right|  $ & Cardinality of a set of objects\\
		$ \omega $ & Convexity status of a cluster after analysis\\
		$ \psi $ & Candidate point for proving the non-convexity of a cluster\\
		\hline
	\end{tabular}
	
\end{table}

\textbf{Definition 1. }(Grid Structure) Let the value space, which is about to be calibrated, consist of $ p $ variables, then a grid structure $ \mathcal{G} $ is a partitioning of the data space, utilizing grid points, into finite number of non-overlapping hypercubic regions, called cells. The extrema of this grid in each dimension are taken as the minimum and maximum values of the related attribute.

\textbf{Definition 2. }(Grid Accuracy) The grid accuracy $ \varepsilon $ defines the size of every cell of $ \mathcal{G} $ in each dimension.

\textbf{Definition 3. }(Neighboring Set of a Grid Point) For the arbitrary grid point $ g = \left( g_1,\cdots,g_i,\cdots,g_p \right) $ in $ \mathcal{G} $, its neighbors are defined as those grid points with the exact distance of $ \varepsilon $ from it. It is like as these neighboring points are lying on a hypersphere, with $ g $ as its centroid and $ \varepsilon $ as its radius. Hence, w.r.t. the grid accuracy, the neighboring grid points of $ g $ in the $ j $th dimension will be $ \left( g_1,\cdots,g_i-\varepsilon,\cdots,g_p \right) $ and $ \left( g_1,\cdots,g_i+\varepsilon,\cdots,g_p \right) $. That is:

\begin{gather*}
\mathbb{N} \left( g \right) = \bigcup_{i=1}^p \left( g_1,\cdots,g_i\pm\varepsilon,\cdots,g_p \right)
\end{gather*}

Therefore, one can state that for every grid point in $ \mathcal{G} $, there are $ 2p $ neighbors\footnote{For grid points at the extrema of the grid, the number of neighbors would be less.}.

\textbf{Definition 4. }($ \varepsilon $ Neighborhood of a Point) For arbitrary point $ x $ in euclidean space, its $ \varepsilon $ neighborhood is a set which contains all arbitrary objects with a distance less than or equal to $ \varepsilon $. That is:

\begin{gather*}
\mathcal{N}_\varepsilon \left( x \right) = \left\lbrace y \mid D \left( x,y \right) \le \varepsilon \right\rbrace, y \in \mathcal{N}_\varepsilon \left( x \right) \iff x \in \mathcal{N}_\varepsilon \left( y \right)
\end{gather*}

where $ D \left( x,y \right) $ denotes the euclidean distance between objects $ x $ and $ y $. Moreover, $ \varepsilon $ vicinity is a symmetric concept.

\section{Proposed Approach}\label{sec_PropMeth}Our proposed approach consists of three major phases. At the first phase, we find those grid points which do not fall at the coverage area by the cluster shape. At the second phase, we utilize such points to detect cluster points which are located at its inner and outer edges. Finally, at the third phase, we evaluate the midpoint convexity on these edging points.

\textbf{Remark 1. }According to the fact that, number of grid points $ t $, increases exponentially w.r.t. the number of dimensions $ p $, hence a situation might happen in which, the amount of grid points is so huge that cannot fit into memory, and also, the consequent computational complexity would be intolerable too. Therefore, in a preprocessing step, one can resort to dimensionality reduction approaches, as in them, the pairwise euclidean distances between data points are approximately preserved. Popular methods are PCA\unskip~\cite{pearson1901liii} and Random Projections (RP)\unskip~\cite{johnson1984extensions,achlioptas2001database}. The main difference between these methods is that PCA is computationally more expensive than RP, but in reverse, its accuracy is more than RP in most cases. Moreover, PCA is more sensitive to the choice of the number of reduced dimensions, while the accuracy for RP increases normally with the number of dimensions, as long as it is desirable for lower dimensions too\unskip~\cite{deegalla2006reducing}.

\textbf{Remark 2. }Sometimes, even by reducing the dimensions of the value space, the amount of grid points is still far high, that would take so much time to be processed. Hence, by losing some precision, in the same preprocessing step, one can conduct a random sampling on grid points, with a reasonable rate, and still expect remarkable results out of the experiments\footnote{Sometimes, in a case that the number of cluster objects is very high, we can carry out a random sampling on data points too, to lower the computational complexity. But this sampling rate should not be so low by which, the structure of the cluster would become so sparse and distorted, and thus, not reliable for being investigated for convexity.}.

\textbf{Remark 3. }For the grid accuracy, we have to define it in an appropriate manner, which would lead to precise detection results. Here, we prefer to use the neighborhood parameter epsilon of DBSCAN algorithm\unskip~\cite{ester1996density}, as the grid accuracy. We consider the optimal value for epsilon, as by which, DBSCAN will report a unique density-based cluster without any noise, as its output. The reason is described as follows. In truth, we can divide the value space containing the cluster into two distinct partitions. One is the space, which is covered by the cluster shape, if we consider it as a distribution with an infinite number of points. The other partition will be the space not including any points of that distribution.

Now, if we intend to establish the grid in a way that the frontiers of the cluster will be defined properly, we should set the grid accuracy in a manner which by that, every grid point which is in the coverage region of the cluster, will be in the $ \varepsilon $ neighborhood of at least one cluster point.

If the grid accuracy is set too low, then there will be some grid points which cannot take place in the $ \varepsilon $ neighborhood of any cluster point, while they are in the covered region by the cluster. In reverse, if it is established too high, then the frontiers of the cluster will not be discovered with desirable precision.

Therefore, we define the accuracy value of the grid equal to the epsilon parameter of DBSCAN, as it is good enough to contain all the grid points in the coverage area by the cluster, and also, to establish the reliable boundaries of the cluster\footnote{With increase in the density of the distribution, the value of $ \varepsilon $ will decrease, and the final precision will increase.}.

The framework of proposed approach is presented in Algorithm~\ref{algo_GridCnvxAnls}, which consists of a preliminary phase and three other major phases including: 0) Initializing the grid; 1) Detecting non-neighboring grid points; 2) Finding marginal cluster points; 3) Analyzing midpoint convexity. All of these phases will be described in details in following subsections.

\begin{algorithm}
	\DontPrintSemicolon
	\SetAlgoLined
	
	\caption{$ \left[ \omega,\psi \right] = $ GridConvxAnals($ \mathcal{X},\varepsilon,\eta $)}
	\label{algo_GridCnvxAnls}
	
	\SetKwInOut{Input}{Input}\SetKwInOut{Output}{Output}
	\Input{$ \mathcal{X} $ - Input cluster; $ \varepsilon $ - Grid accuracy; $ \eta $ - Random sampling rate for grid points}
	\Output{$ \omega $ - Convexity status; $ \psi $ - Candidate point for non-convexity}
	\BlankLine
	
	\textbf{Phase 0 {\textemdash} Initializing the grid:}\BlankLine
	\textit{Step 1.} Add a distance of $ 2\varepsilon $ to the extrema of the value space in each dimension, and initialize the grid structure $ \mathcal{G} $, regarding the grid accuracy $ \varepsilon $\;
	
	\textit{Step 2.} Create $ \eta t $ number of grid points totally at random, w.r.t. $ \mathcal{G} $, and assign them to $ \mathcal{G}_s $\;
	
	\BlankLine
	
	\textbf{Phase 1 {\textemdash} Detecting non-neighboring grid points:}\BlankLine
	\textit{Step 3.} For every grid point in $ \mathcal{G}_s $ evaluate whether or not it falls in the $ \varepsilon $ neighborhood of any cluster point, and separate the non-neighboring grid points as a distinct set, all w.r.t. Algorithm~\ref{algo_detcNonNghbGrdPnts}\;
	
	\textbf{Phase 2 {\textemdash} Finding marginal cluster points:}\BlankLine
	\textit{Step 4.} According to Algorithm~\ref{algo_detcMargClstPnts}, regarding the set of non-neighboring grid points, find those neighbors of these points, which are at the $ \varepsilon $ neighborhood of at least on cluster point. Consider the nearest cluster point as one of the frontier points of the cluster\;
	
	\textbf{Phase 3 {\textemdash} Analyzing midpoint convexity:}\BlankLine
	\textit{Step 5.} With respect to Algorithm~\ref{algo_midPntCnvxAnls}, for every distinct pair of frontier points of the cluster, find the related midpoint, and analyze whether this midpoint falls in the $ \varepsilon $ neighborhood of any cluster point. If it is not so, then announce that the cluster is non-convex, and provide that midpoint as a candidate point for proving non-convexity of the cluster\;
	
\end{algorithm}

\setcounter{subsection}{-1}
\subsection{Initializing the grid}\label{subSec_InitGrid}At this early stage, we establish the grid structure which is covering the value space of the cluster. For this matter we should divide each dimension into smaller pieces equal to $ \varepsilon $ in size. But as for some extreme grid points which might be at the $ \varepsilon $ neighborhood of a cluster object, and thus, the related marginal cluster points will not be capable of being found in our process, therefore, we add a distance of $ 2\varepsilon $ to the extrema of each dimension of the value space. Hence, even at the extreme regions of a cluster in any dimension, there will be at least one grid point which is not covered by the cluster shape.

After creating the grid vectors, by simple randomness, we choose a value in each vector and thus, create an arbitrary grid object. As we are going to work with a portion of total grid points, we conduct the process multiple times and create the set of sampled grid points. In the following, this set will represent the whole grid with some precision proportional to the random sampling rate $ \eta $ for grid structure.

\subsection{Detecting Non-Neighboring Grid Points}\label{subSec_DetcNonNghbGridPnts}After attaining the sampled grid, we afford to find those grid objects which do not fall in the $ \varepsilon $ neighborhood of any cluster object. In other words, such points are not covered by the cluster shape. The reason that at the first step, we are trying to locate these points is that by utilizing them, we can find those grid neighbors of them, which are at the boundaries of the cluster. Algorithm~\ref{algo_detcNonNghbGrdPnts} demonstrates the required steps for finding non-neighboring grid points.

\begin{algorithm}
	\DontPrintSemicolon
	\SetAlgoLined
	
	\caption{$ \left[ \mathcal{T} \right] = $ DetcNonNghbGridPnts($ \mathcal{X},\mathcal{G}_s,\varepsilon $)}
	\label{algo_detcNonNghbGrdPnts}
	
	\SetKwInOut{Input}{Input}\SetKwInOut{Output}{Output}
	\Input{$ \mathcal{X} $ - Input cluster; $ \mathcal{G}_s $ - Set of sampled grid points; $ \varepsilon $ - Grid accuracy}
	\Output{$ \mathcal{T} $ - Set of non-neighboring grid points}
	\BlankLine	
	
	$ \mathcal{T} \leftarrow \Phi $\;
	\ForEach{$ g \in \mathcal{G}_s $}{
		$ \zeta \leftarrow 0 $\;
		\ForEach{$ x \in \mathcal{X} $}{
			\If{$ g \in \mathcal{N}_\varepsilon \left( x \right) $}{
				$ \zeta \leftarrow 1 $\;
				break\;
			}
		}
		\If{$ \zeta \equiv 0 $}{
			$ \mathcal{T} \leftarrow \mathcal{T} \cup g $\;
		}
	}
	
\end{algorithm}

\subsection{Finding Marginal Cluster Points}\label{subSec_FindMargClstPnts}By gaining the non-neighboring grid points, as it was mentioned earlier, we afford to find those neighbors of such grid objects, which fall at the $ \varepsilon $ neighborhood of at least one cluster object\footnote{If the whole grid is utilized, all of the neighbors of non-neighboring grid points, excluding the non-neighboring objects themselves, will be the same marginal grid points. And we just need to find the closest cluster object to each of them.}. By finding that closest cluster object, we mark it as one of the bordering points of the cluster shape. Algorithm~\ref{algo_detcMargClstPnts}, shows the required steps.

\begin{algorithm}
	\DontPrintSemicolon
	\SetAlgoLined
	
	\caption{$ \left[ \mathcal{V} \right] = $ DetcMargClstPnts($ \mathcal{X},\mathcal{T},\varepsilon $)}
	\label{algo_detcMargClstPnts}
	
	\SetKwInOut{Input}{Input}\SetKwInOut{Output}{Output}
	\Input{$ \mathcal{X} $ - Input cluster; $ \mathcal{T} $ - Set of non-neighboring grid points; $ \varepsilon $ - Grid accuracy}
	\Output{$ \mathcal{V} $ - Set of marginal cluster points}
	\BlankLine	
	
	$ \mathcal{U} \leftarrow \Phi $\;
	$ \mathcal{V} \leftarrow \Phi $\;
	\ForEach{$ h \in \mathcal{T} $}{
		$ \mathcal{U} \leftarrow \mathcal{U} \cup \mathbb{N}\left( h\right) $\;
	}
	$ \mathcal{U} \leftarrow \mathcal{U} \setminus \mathcal{T} $\;
	\ForEach{$ i \in \mathcal{U} $}{
		$ \mathcal{W} \leftarrow \Phi $\;
		\ForEach{$ x \in \mathcal{X} $}{
			\If{$ i \in \mathcal{N}_\varepsilon \left( x \right) $}{
				$ \mathcal{W} \leftarrow \mathcal{W} \cup x $\;
			}
		}
		\If{$ \left| \mathcal{W} \right| \neq 0 $}{
			$ \mathcal{V} \leftarrow \mathcal{V} \cup \argmin_{j\in \mathcal{W}} D \left( i,j \right) $\;
		}
	}
	
\end{algorithm}

\subsection{Analyzing Midpoint Convexity}\label{subSec_AnlsMdptCnvx}After building the approximate boundaries of the cluster structure, it is time to analyze the notion of midpoint convexity on the marginal cluster points. Therefore, we evaluate every distinct pair of such points, whether or not their midpoint falls in the $ \varepsilon $ neighborhood of at least one cluster object. If there is at least one pair of frontier points, which their midpoint is not covered by the cluster shape, hence, the cluster will be reported as non-convex. Otherwise, it would be convex with the precision of $ \varepsilon $. Algorithm~\ref{algo_midPntCnvxAnls}, illustrates the needed actions to evaluate midpoint convexity on frontier objects of the cluster.

\begin{algorithm}
	\DontPrintSemicolon
	\SetAlgoLined
	
	\caption{$ \left[ \omega,\psi \right] = $ MidPntCnvxAnls($ \mathcal{X},\mathcal{V},\varepsilon $)}
	\label{algo_midPntCnvxAnls}
	
	\SetKwInOut{Input}{Input}\SetKwInOut{Output}{Output}
	\Input{$ \mathcal{X} $ - Input cluster; $ \mathcal{V} $ - Set of marginal cluster points; $ \varepsilon $ - Grid accuracy}
	\Output{$ \omega $ - Convexity status; $ \psi $ - Candidate point for non-convexity}
	\BlankLine	
	
	$ \omega \leftarrow 1 $\;
	$ \psi \leftarrow \phi $\;
	\ForEach{$ \left( j,k \right) \mid j,k \in \mathcal{V} $ and $ j \neq k $}{
		$ \gamma \leftarrow M \left( j,k \right) $\;
		$ \zeta \leftarrow 0 $\;
		\ForEach{$ x \in \mathcal{X} $}{
			\If{$ \gamma \in \mathcal{N}_\varepsilon \left( x \right) $}{
				$ \zeta \leftarrow 1 $\;
				break\;
			}
		}
		\If{$ \zeta \equiv 0 $}{
			$ \omega \leftarrow 0 $\;
			$ \psi \leftarrow \gamma $\;
			break\;
		}
	}
	
\end{algorithm}

\begin{figure*}[t!]
	\centering
	\includegraphics[width=1\linewidth]{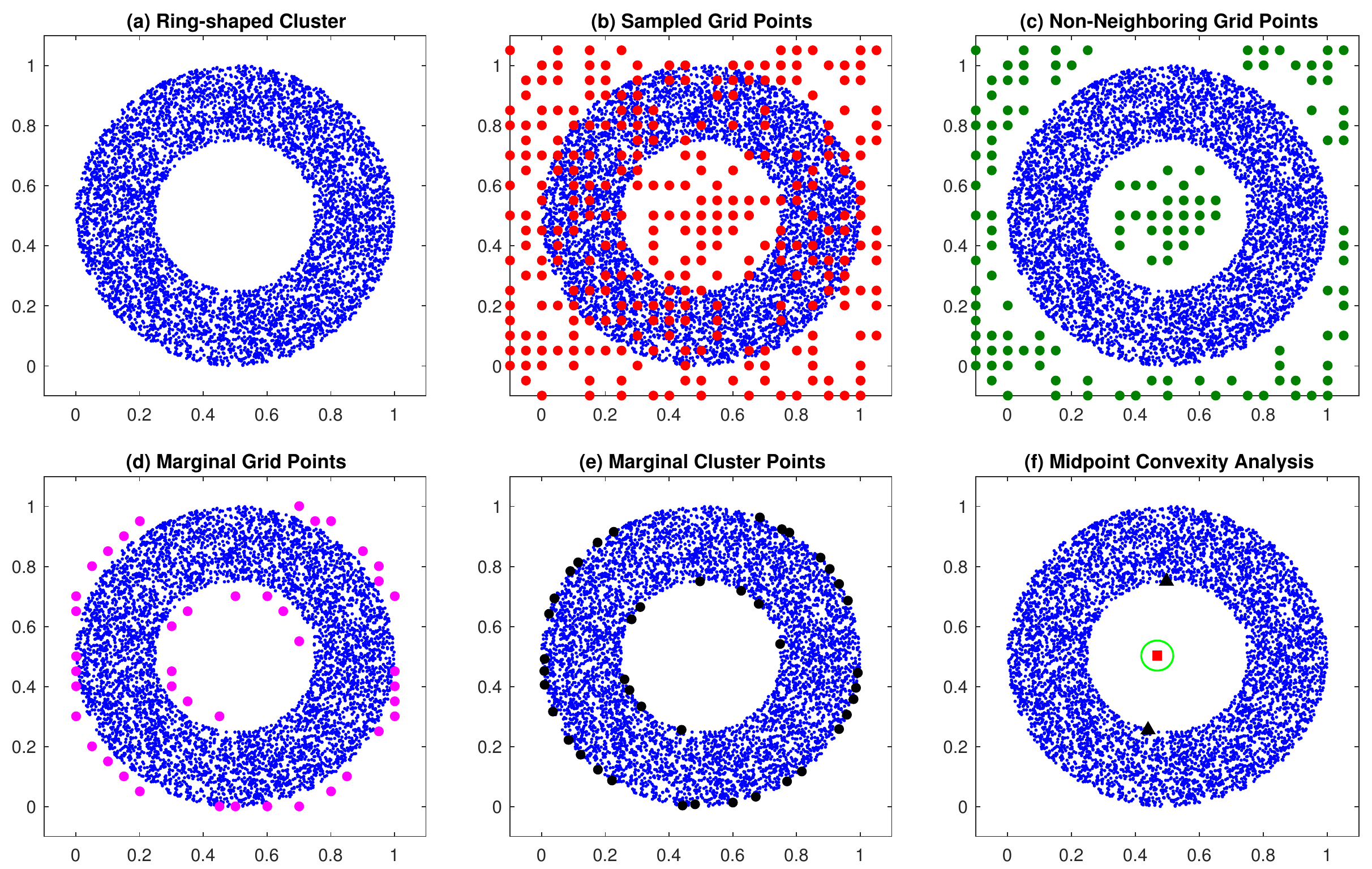}
	\caption{Convexity analysis of a 2D ring-shaped cluster}
	\label{figure-effcResl}
\end{figure*}

\section{Experiments}\label{sec_Expr}In this section, we conduct an efficacy test, to show that our algorithm is capable of detecting the reliable rough margins of a non-convex dense cluster, and proves its non-convexity in a geometrical manner. Moreover, a test is carried out to illustrate the serious dependence of the proposed method on the grid accuracy. Another test is presented to show that even by low rates of random sampling, one can still expect significant detection results out of our approach. All implementations are carried out with MATLAB 9, and run on a laptop with Intel Core i5 processor (clocked at 2.5 GHz) and 6 G memory. Moreover, for reproducibility, we publish our code on GitHub\footnote{\url{https://github.com/sana33/Researches/tree/master/GbACADC}}.

\subsection{Efficacy Test}\label{subSec_EffcTest}While there is not any real-life benchmark data which claims on the convexity of its contained clusters, thus, we run evaluations on a synthetic two-dimensional dataset, with uniform distribution, including only one non-convex cluster in the shape of a ring.

Figure~\ref{figure-effcResl}a demonstrates the ring-shaped 2D cluster, denoted as blue dots, with $ \varepsilon = 0.05 $, as the grid accuracy, as by which, the DBSCAN algorithm will report the related cluster, as a dense shape and without any noise. Figure~\ref{figure-effcResl}b demonstrates the cluster objects along with the sampled grid points shown with red circles, created through the phase of initializing the grid structure, with random sampling rate $ \eta = 0.5 $. Figure~\ref{figure-effcResl}c represents the non-neighboring grid points with green circles, and figure~\ref{figure-effcResl}d illustrates the marginal grid points in magenta circles, which are at the neighborhood of at least one non-neighboring grid object.

Moreover, figure~\ref{figure-effcResl}e demonstrates the frontier cluster objects, denoted as black circles, obtained with the precision of $ \varepsilon $, and as it is clear, both outer and inner frontiers are detected. Finally, figure~\ref{figure-effcResl}f shows a pair of marginal cluster objects denoted as black triangles, and their midpoint denoted as a red square, with its $ \varepsilon $ vicinity denoted as a green circle. It is crystal clear that the midpoint is not in the $ \varepsilon $ neighborhood of any cluster object. Therefore, the dense ring-shaped cluster is reported as a non-convex shape.

\begin{figure}[t]
	\centering
	\includegraphics[width=1\linewidth]{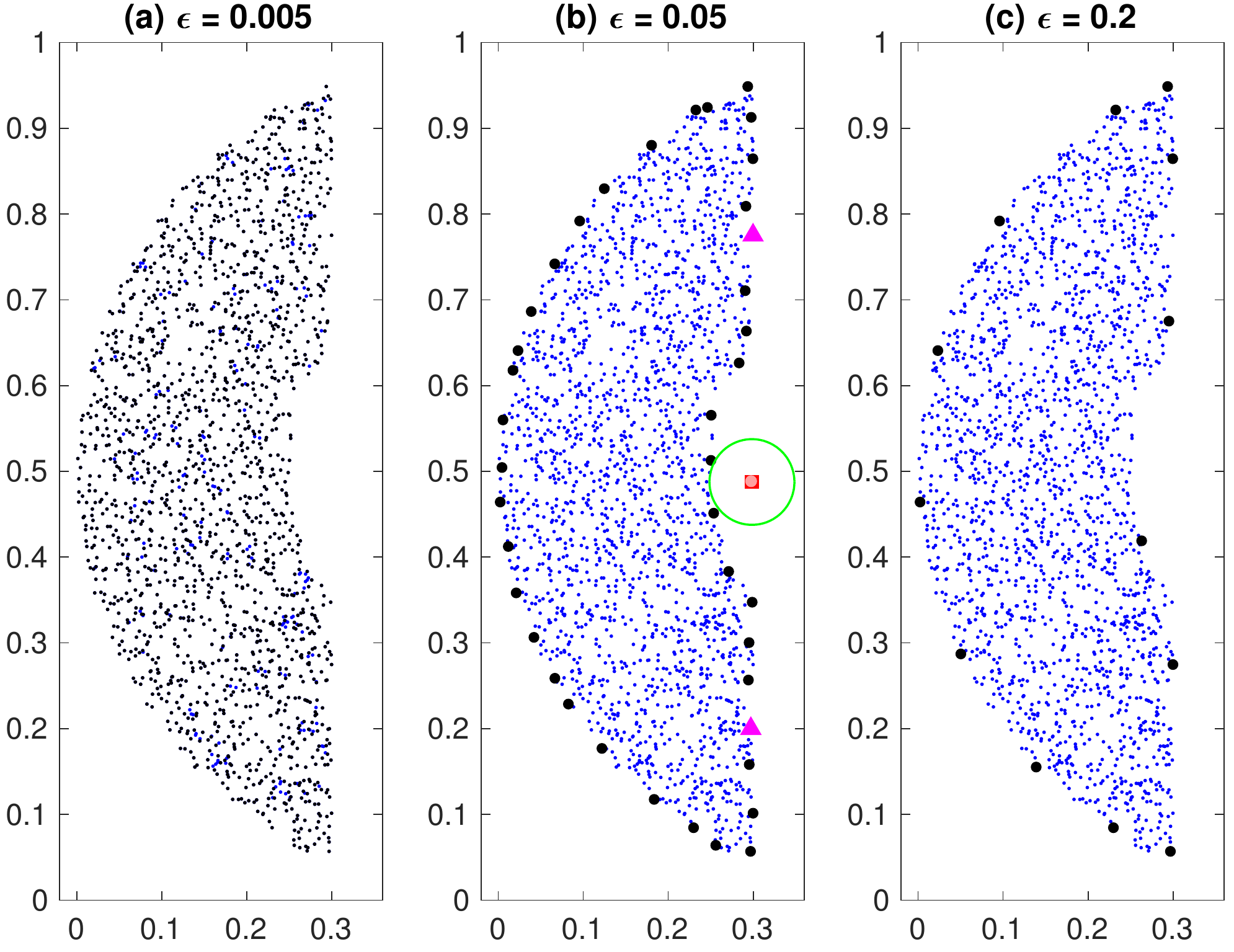}
	\caption{Effect of grid accuracy on detection result}
	\label{figure-testGridAccr}
\end{figure}

\subsection{Test on Grid Accuracy}\label{subSec_testGridAccr}At this part, we demonstrate that if the grid accuracy is not established in a proper manner, then not only the non-convexity might not be discovered, but even a convex-shaped cluster could be declared as non-convex. As it was explained earlier, if the grid accuracy is set too low, then there will be some grid points, erroneously reported as non-neighboring objects. And if it is set too high, the consequent boundaries will not be reported appropriately. Figure~\ref{figure-testGridAccr} illustrates a crescent-like density-based cluster, denoted as blue dots, in three different conditions made by various values for grid accuracy.

Figure~\ref{figure-testGridAccr}a shows the crescent along with the marginal cluster points, denoted as black dots, obtained through $ \varepsilon = 0.005 $. As it is clear, not only the boundaries of the crescent are not detected correctly, but even there are numerous cluster points mistaken as frontier objects. Figure~\ref{figure-testGridAccr}b demonstrates the best result achieved through $ \varepsilon = 0.05 $. Two arbitrary marginal cluster objects and their midpoint, along with its $ \varepsilon $ vicinity, are represented with magenta triangles and a red square and a green circle, respectively. As it is apparent, the midpoint is out of the contained area by the crescent, and hence, convexity will be reported. Figure~\ref{figure-testGridAccr}c shows the crescent along with the boundary objects obtained through $ \varepsilon = 0.2 $. As it is obvious, the related boundaries are not precisely detected, thus, there is not any distinct pair of marginal cluster points with a midpoint out of the coverage region by the cluster, w.r.t. the elected high value for grid accuracy.

\begin{figure}[t]
	\centering
	\includegraphics[width=1\linewidth]{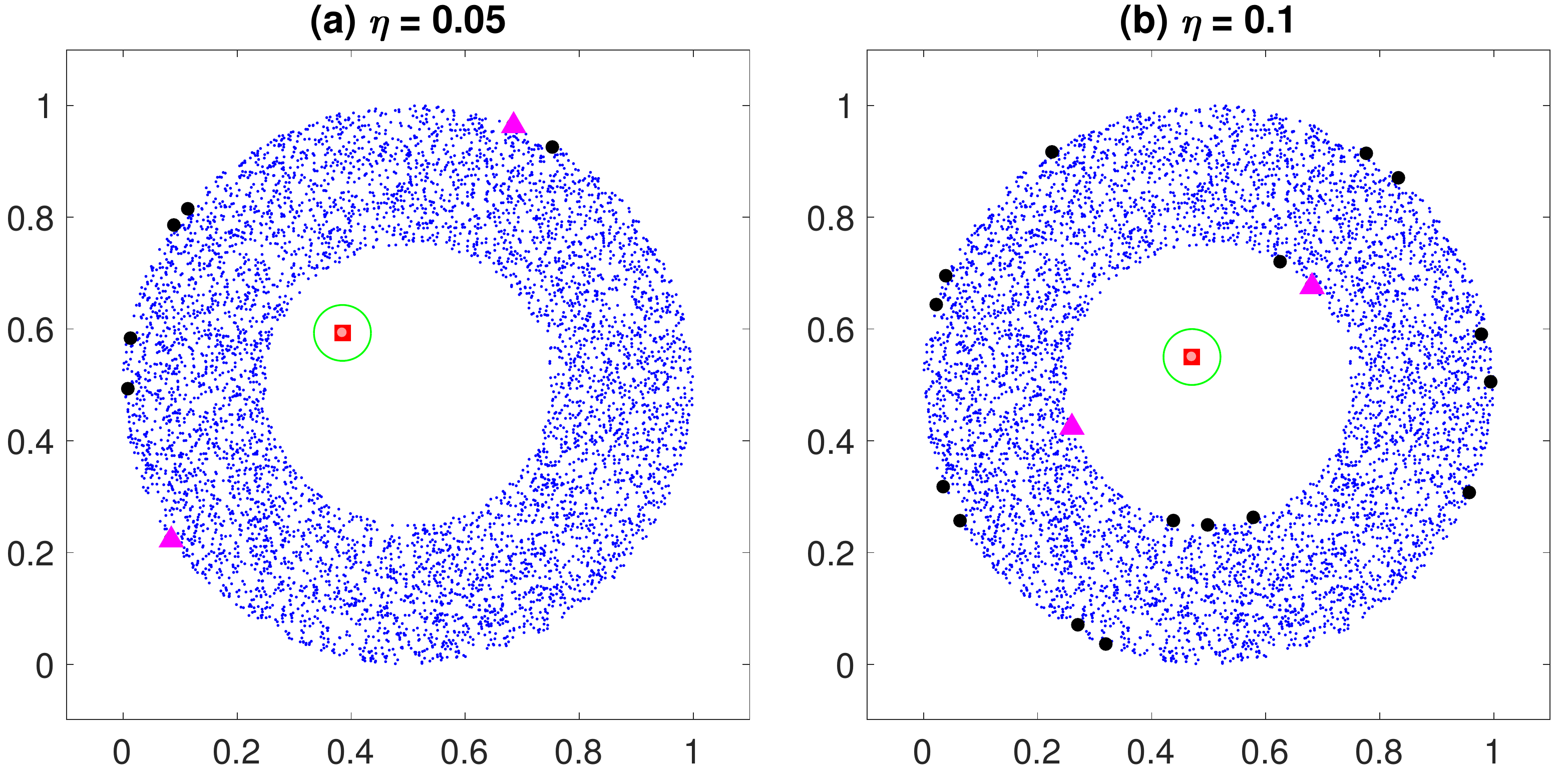}
	\caption{Effect of random sampling rate for the grid, on detection result}
	\label{figure-testRandSamp}
\end{figure}

\subsection{Test on Sampling Rate for the Grid}\label{subSec_testGridSampRate}However, there is no guarantee that with very low rates of random sampling for the grid, one can still anticipate remarkable detection results out of our proposed approach, we illustrate here with two examples run with different very low values of $ \eta $, that it could be possible to discover the non-convexity, even with very small samples of the grid.

As the random sampling rate decreases, the chance at which both inner and outer frontier objects could be detected correctly, reduces monotonically. Figure~\ref{figure-testRandSamp} demonstrates the same ring from efficacy test, with the same graphical representations as in figure~\ref{figure-testGridAccr}. Figure~\ref{figure-testRandSamp}a shows the ring and the related marginal cluster points obtained through $ \eta = 0.05 $. It is clear to see that not any points of the inner margin are detected. But as the position of the inner hole is at the exact center of the ring, there is a pair of outer marginal cluster points that their midpoint falls in the hole, far from the coverage region by the ring. Hence, the ring is declared as a non-convex density-based cluster.

Figure~\ref{figure-testRandSamp}b demonstrates the result achieved out of $ \eta = 0.1 $. However, both inner and outer boundaries of the shape are detected with a very low precision, the non-convexity of the ring is elicited successfully. Therefore, one can state that for better and more accurate detection outcomes, we need to utilize large enough samples of the grid, otherwise the outcome might be unreliable.

\section{Conclusion}\label{sec_Conc}In this paper, we just provided a new plain approach, risen from the field of computational geometry, to examine the convexity of a density-based cluster. For this matter, first of all we initialized a grid and tried to locate those grid objects, which are not covered by the cluster shape. By utilizing these objects, we afforded to find marginal cluster objects, and built the approximate inner and outer frontiers of the cluster upon them. Finally, by employing the concept of midpoint convexity on these boundary points, we discovered whether the cluster shape is convex.





%

\bibliographystyle{IEEEtran}

\bibliography{\jobname}

\end{document}